\documentclass[]{article}

\usepackage{PRIMEarxiv}

\usepackage[utf8]{inputenc} 
\usepackage[T1]{fontenc}    
\usepackage{hyperref}       
\usepackage{url}            
\usepackage{booktabs}       
\usepackage{amsfonts}       
\usepackage{nicefrac}       
\usepackage{microtype}      
\usepackage{lipsum}
\usepackage{fancyhdr}       
\usepackage{graphicx}       
\graphicspath{{media/}}     
\usepackage{wrapfig}
\usepackage{amsmath,amssymb,amsfonts}
\usepackage[flushleft]{threeparttable}

\title{PyTorch Image Quality: \\ Metrics for Image Quality Assessment
}

\author{
  Sergey Kastryulin \\
  Computational Imaging Laboratory \\
  Skolkovo Institute of Science and Technology \\
  Moscow, Russia \\
  \texttt{sergey.kastryulin@skoltech.ru} \\
   \And
  Jamil Zakirov \\
  Independent Researcher \\
  Skolkovo Institute of Science and Technology \\
  Moscow, Russia\\
  \texttt{jamil.zakirov@skoltech.ru} \\
  \And
  Denis Prokopenko \\
  Biomedical Engineering Department \\
  School of Biomedical Engineering and Imaging Sciences \\
  King's College London, London, UK\\
  \texttt{d.prokopenko@outlook.com} \\
  \And
  Dmitry V. Dylov \\
  Computational Imaging Laboratory \\
  Skolkovo Institute of Science and Technology \\
  Moscow, Russia \\
  \texttt{d.dylov@skoltech.ru} \\
}

\begin{document}
\maketitle

\begin{abstract}

Image Quality Assessment (IQA) metrics are widely used to quantitatively estimate the extent of image degradation following some forming, restoring, transforming, or enhancing algorithms.
We present PyTorch Image Quality (PIQ), a usability-centric library that contains the most popular modern IQA algorithms, guaranteed to be correctly implemented according to their original propositions and thoroughly verified.
In this paper, we detail the principles behind the foundation of the library, describe the evaluation strategy that makes it reliable, provide the benchmarks that showcase the performance--time trade-offs, and underline the benefits of GPU acceleration given the library is used within the PyTorch backend. 
PyTorch Image Quality is an open source software: \href{https://github.com/photosynthesis-team/piq}{https://github.com/photosynthesis-team/piq/}.

\end{abstract}


\section{Introduction}


\begin{figure*}
    \centering
    \includegraphics[width=0.8\linewidth]{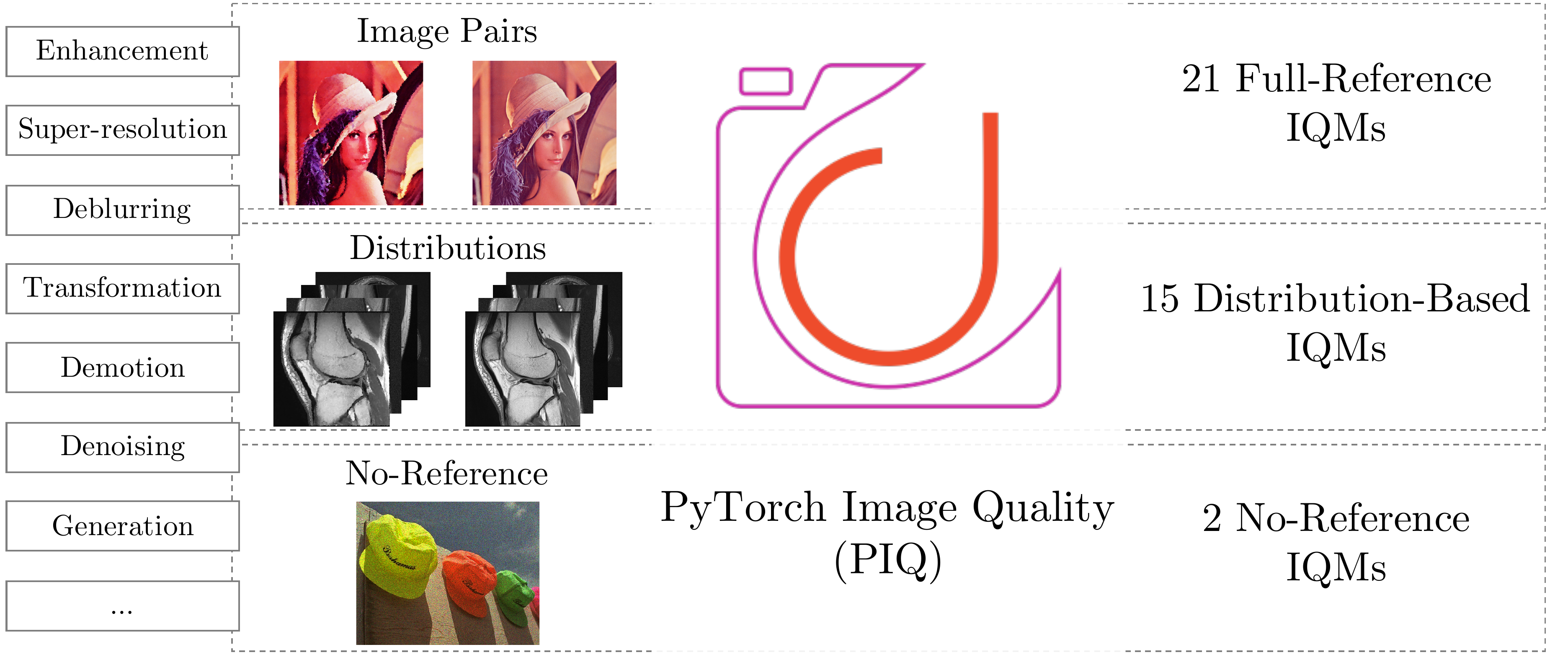}
    \caption{PIQ open source library implements 38+ metrics in three main categories: Full-Reference (FR), No-Reference (NR), and Distribution-Based (DB). 
    Our implementations enable quantitative estimation of the quality of images from various domains, including medical scans.
    Right column shows the counters of the currently implemented metrics.
    Note that DB metrics can be used with different feature extractors, further diversifying the pool of available metrics.}
    \label{fig:salesfig}
\end{figure*}

With the ever-rising interest towards generation, recovery, and enhancement of images, a number of computational Image Quality Assessment (IQA) methods have become available. 
These methods aim to produce a score that would perfectly correlate with the human perception of the Image Quality (IQ). 
Given the large number of the assorted IQA options and frequent mismatch between different implementations of the same metric, a demand for reliable and community-validated resource has become evident.
Specifically, such a resource should embrace a plethora of possible modern image metrics, eliminating the need to search, install, or re-implement the algorithms.

Although popular IQA libraries provide the access to some Image Quality Metrics (IQMs) in the form of user-friendly packages, they do have a set of limitations.
Blind Image Quality Toolbox \cite{sollinger2017blind_iqa_toolbox} and Image Quality Assessment Toolbox \cite{xing2021iqa_toolbox} are verified MATLAB implementations of no-reference (NR) and full-reference (FR) IQMs. 
Unfortunately, the growing popularity of Python makes these solutions less relevant for the hands-on research efforts, where the majority of image and vision computing experts rely on Python for its well-tested open source repositories.
Kornia \cite{eriba2019kornia}, PIQA \cite{rozet2020piqa}, and IQA-PyTorch \cite{chen2020iqa_pytorch} do provide Python interfaces for the metrics and the IQA-optimization \cite{ding2020iqa_optimization}, while also focusing on their use as the loss functions. 
However, these libraries implement a rather limited number of algorithms\footnote{Number of IQMs implemented: Kornia - 2, PIQA - 11, IQA-PyTorch - 25.}.

This manuscript introduces PyTorch Image Quality (PIQ) open source library that provides an extended collection of implementations of IQA metrics and the corresponding loss functions, using PyTorch \cite{pytorch} to enable fast and efficient computations on the graphics processing unit (GPU). 
This work is the result of three years of collecting and comparing the best image quality metrics in one place, with independent implementation, optimization, and testing. 
At the time of writing this manuscript, our GitHub project receives over 4000 monthly views, having accumulated 640 stars and having been forked over 57 open-source projects. 
This tool has been actively used in the research community, including the development of new neural network architectures \cite{lunardi2022arcade} and the large-scale medical image quality study \cite{kastryulin2022iqa_mri}.

\section{Design principles}

We created PyTorch Image Quality to maximise the benefit for the research community working on a plethora of image-to-image translation problems in a variety of visual data domains (Fig.~\ref{fig:salesfig}). 
To do that, we build the library based on the following design principles:

\textbf{Be user-friendly}. \space Unfortunately, the complexity inherent to the field of Computer Vision is multiplied by that of the sophisticated algorithms designed to boost the performance of the IQA approaches. 
We hide this internal complexity behind easy-to-use APIs that follow the principle of the least astonishment \cite{smith2015good_library} to provide a seamless user experience.

\textbf{Be reliable}. \space Today, publicly available implementations may produce inconsistent results, which applies even to the most well-known and the widely used IQMs (\textit{e.g.} SSIM \cite{wang2004image}).
PyTorch Image Quality library is focused on a thorough testing to provide consistency with the formal metrics definitions and the original implementations proposed by their authors (if these implementations exist).

\textbf{Be pragmatic}. \space A majority of modern (and some classical) IQMs are computationally inefficient by their design, which hinders the ultimate performance.
Therefore, PyTorch Image Quality purposely enables an optimization when its extra complexity is worth delivering a compelling performance.
Being inspired by the same principle in PyTorch, we state that trading 10\% of speed for a model that is significantly easier to use is acceptable; and 100\% is not \cite{pytorch}.
\section{Usability--centric approach}

Currently, PyTorch Image Quality library contains implementations of the following 38 metrics: \textbf{21 Full-Reference (FR) IQMs} (PSNR, SSIM \cite{wang2004image}, MS-SSIM \cite{wang2003multiscale}, IW-SSIM \cite{wang2010information}, VIF \cite{sheikh2005visual}, GMSD \cite{xue2013gradient},  MS-GMSD and MS-GMSDc \cite{Zhang2017gmsd}, FSIM and FSIMc \cite{zhang2011fsim}, SR-SIM and SR-SIMc \cite{zhang2012srsim}, VSI \cite{zhang2014vsi}, MDSI \cite{nafchi2016mdsi}, HaarPSI \cite{Reisenhofer2018haarpsi}, Content and Style Perceptual Scores \cite{johnson2016perceptual}, LPIPS \cite{zhang2018unreasonable}, DISTS \cite{ding2020image}, PieAPP \cite{prashnani2018pieapp}, DSS \cite{Balanov2015dss}), 
\textbf{2 No-Reference (NR) IQMs }(BRISQUE \cite{MittalMB12}, Total Variation),
and \textbf{15 Distribution-Based (DB) IQMs} (KID \cite{binkowski2018demystifying}, FID \cite{heusel2017gans}, GS \cite{khrulkov2018geometry}, Inception Score (IS) \cite{salimans2016improved}, MSID \cite{tsitsulin2020shape}, all implemented with three different feature extractors: Inception Net \cite{szegedy2015going}, VGG16, and VGG19 \cite{simonyan2014very}).
The general taxonomy of IQA metrics, their detailed description, and a pertinent discussion can be found in \ref{sec:appendix}.

\textbf{\textit{Remark}}. Throughout the work, we use the term ``\textit{metrics}'' to describe the IQA algorithms.
Technically, this term is not mathematically correct, because a \textit{metric} is a function for which the identity of indiscernibles, the symmetry, and the triangle inequality must hold \cite{choudhary1993metric_spaces}.
For the majority of implemented algorithms, one or several of these attributes do not hold.
However, the term is still used here, which reflects the commonplace convention in the community.

\subsection*{Metrics as loss functions}

Previous studies of IQMs showed their efficiency in reflecting the human perception of visual quality \cite{athar2019comprehensive, Mason2020medicalcomparison, ding2020comparison, kastryulin2022iqa_mri}.
That brings a temptation to use the best-performing differentiable IQMs as the loss functions for a direct optimization of the image processing models \cite{ding2021metrics_as_losses}.
To enable that possibility, all PIQ metrics are well-integrated with the PyTorch \cite{pytorch} backend, enabling the automatic computing of the gradients of the differentiable models.

Besides, the use of PyTorch enables the GPU acceleration, yielding a faster computation of the metrics and the losses (see Fig. \ref{fig:complexity-performance-gpu} for more details).
Moreover, our implementation strategy allows a seamless integration with the most common deep learning pipelines in PyTorch.
For instance, the flexible interface enables our metric implementations to be used as additional layers of a deep neural network.

\subsection*{Integrated feature extractors}

DB IQMs are computed on features obtained from images with feature extractors, which are typically represented by pre-trained convolutional neural networks. 
Most methods are evaluated using a single feature extractor (typically, Inception Net \cite{szegedy2015going}).
A recent study on IQMs for Magnetic Resonance Imaging \cite{kastryulin2022iqa_mri} showed that the choice of feature extractor plays a critical role, heavily influencing the performance of DB metrics.
These result may inspire our users to experiment with various feature extractor options.
To address that, we added a possibility to provide their own feature extractors or choose one of the models integrated into the library (Inception Net \cite{szegedy2015going}, VGG16, and VGG19 \cite{simonyan2014very}).

\subsection*{Chromatic versions of luminance based metrics}

The majority of methods are designed to be applied to images in RGB color space.
However, some of them (\textit{e.g.} FSIM \cite{zhang2011fsim} and SR-SIM \cite{zhang2012srsim}) are designed for grayscale images or for the luminance component of the color images.
Because the chrominance information also affects human visual system (HVS) in understanding the images, better performance can be expected if the chrominance information is incorporated for color IQA. 
For that, we follow the common approach of first converting the RGB images into the YIQ colour space. 
The Y channel is used for the computations of the initial grayscale variants, while the I and the Q components are added to obtain the chromatic versions of the metrics (\textit{e.g.} FSIMc \cite{zhang2011fsim} and SR-SIMc \cite{zhang2012srsim} respectively). 

\section{Evaluation}

New metrics and measures for IQA that claim to be better than their predecessors emerge every day.
To prove that, authors show IQMs' performance on specifically designed IQA datasets consisting of pairs of distorted and reference images accompanied with similarity scores estimated by human assessors.
The availability of such datasets allows to compute correlations between human scores and metrics' estimates of quality to find algorithms that best reflect judgement of HVS.

\begin{table*}[tbh!]
\caption{Overview of existing IQA datasets}
\vspace{3mm}
\setlength\tabcolsep{2.5pt}
\begin{center}
\resizebox{.8\columnwidth}{!}{ 
\begin{tabular}{c|cccc}
\toprule

Database &Year & Reference & Distorted & Subjective Score\\
\midrule

IVC \cite{le2005subjective}                     & 2005 & 10      & 235     & MOS  (1 $\sim$ 5)        \\
LIVE IQA \cite{sheikh2006statistical, LIVE_R2}  & 2006 & 29      & 779     & DMOS (0 $\sim$ 100)      \\ 
A57 \cite{chandler2007a57, chandler2007vsnr}    & 2007 & 3       & 54      & DMOS (0 $\sim$ 1)        \\
Toyama/MICT \cite{horita2011mict}               & 2008 & 14      & 168     & MOS  (1 $\sim$ 5)        \\
TID2008 \cite{ponomarenko2009tid2008}           & 2008 & 25      & 1,700   & MOS  (0 $\sim$ 9)        \\
CSIQ \cite{larson2010most, CSIQ}                & 2009 & 30      & 866     & DMOS (0 $\sim$ 1)        \\
IVC-LAR \cite{Strauss2009ivclar}                & 2009 & 8       & 120     & MOS  (1 $\sim$ 5)        \\
WIQ \cite{2009wiq_dataset}                      & 2009 & 7       & 80      & DMOS (0 $\sim$ 100)      \\
IRSQ \cite{Ma2012irsq}                          & 2012 & 57      & 171     & MOS  (0 $\sim$ 5)        \\
VCLFER \cite{zaric2012vcl, VCLFER}              & 2012 & 23      & 552     & MOS  (0 $\sim$ 100)      \\
LIVE MD \cite{jayaraman2012objective, LIVE_MD}  & 2013 & 15      & 405     & DMOS (0 $\sim$ 100)      \\
TID2013 \cite{ponomarenko2015image}             & 2013 & 25      & 3,000   & MOS  (0 $\sim$ 9)        \\
MDID2013 \cite{gu2014hybrid}                    & 2014 & 12      & 324     & DMOS (0.3 $\sim$ 0.6)    \\
CID2013 \cite{virtanen2015cid2013}              & 2013 & 8       & 480     & MOS  (0 $\sim$ 9)        \\
CIDIQ \cite{Liu2014CIDIQ}                       & 2014 & 23      & 690     & MOS  (0 $\sim$ 9)        \\
SIQAD \cite{yang2015perceptual}                 & 2015 & 20      & 980     & DMOS (0 $\sim$ 100)      \\
LIVE in the wild (CLIVE) \cite{Ghadiyaram2016livewild1, 
Ghadiyaram2016livewild2}                        & 2015 & -       & 1,162   & MOS  (1 $\sim$ 5)        \\
MD-IVL \cite{corchs2017multidistortion, MD-IVL} & 2017 & 10      & 750     & MOS  (0 $\sim$ 100)      \\
MDID \cite{sun2017mdid}                         & 2017 & 20      & 1,600   & MOS  (0 $\sim$ 9)        \\
KonIQ-10k \cite{hosu2020koniq}                  & 2018 & 10,073  &  10,073 & MOS  (1 $\sim$ 100)      \\
KADID-10k \cite{kadid10k}                       & 2019 & 81      & 10,125  & DMOS (1 $\sim$ 5)        \\
KADIS-700k \cite{lin2020deepfl}                 & 2019 & 140,000 & 700,000 & DMOS (1 $\sim$ 5)        \\
PaQ-2-PiQ \cite{ying2019paqtopiq}               & 2019 & -       & 40,000  & MOS  (0 $\sim$ 100)      \\
SPAQ \cite{fang2020spaq}                        & 2020 & 11,125  & -       & MOS  (0 $\sim$ 100)      \\
PIPAL \cite{pipal, Gu2021ntire, gu2020pipal}    & 2020 & 250     & 29,000  & MOS  (Elo rating system) \\
\bottomrule
\end{tabular}
}
\end{center}
\label{table:iqa_datasets}
\end{table*}

We use these results to evaluate the correctness of our implementations.
For that, implemented metrics are computed on selected IQA datasets and obtained values are compared with the ones reported in corresponding research papers.

\begin{table*}[tbh!]
\caption{Comparison of correlation values (in terms of SRCC) reported in literature with PIQ implementations.}
\vspace{3mm}
\setlength\tabcolsep{2.5pt}
\begin{center}
\resizebox{.7\columnwidth}{!}{ 
\begin{tabular}{llllll}
\toprule

&\multicolumn{1}{c}{TID2013}&&\multicolumn{1}{c}{KADID-10k}&&\multicolumn{1}{c}{PIPAL}  \\
 \cmidrule{2-2} \cmidrule{4-4} \cmidrule{6-6}
 & PIQ / Reference && PIQ / Reference && PIQ / Reference \\
  \cmidrule{1-6}
  PSNR                                  & $0.69$ / $0.69$ \cite{ponomarenko2015image}   && $0.68$ / $-$                    && $0.41$ / $0.41$ \cite{pipal} \\
  SSIM \cite{wang2004image}             & $0.72$ / $0.64$ \cite{ponomarenko2015image}   && $0.72$ / $0.72$ \cite{kadid10k} && $0.50$ / $0.53$ \cite{pipal} \\
  MS-SSIM \cite{wang2003multiscale}     & $0.80$ / $0.79$ \cite{ponomarenko2015image}   && $0.80$ / $0.80$ \cite{kadid10k} && $0.55$ / $0.46$ \cite{pipal} \\
  IW-SSIM \cite{wang2010information}    & $0.78$ / $0.78$ \cite{athar2019comprehensive} && $0.85$ / $0.85$ \cite{kadid10k} && $0.60$ / $-$                 \\
  VIFp \cite{sheikh2005visual}          & $0.61$ / $0.61$ \cite{ponomarenko2015image}   && $0.65$ / $0.65$ \cite{kadid10k} && $0.50$ / $-$                 \\
  GMSD \cite{xue2013gradient}           & $0.80$ / $0.80$ \cite{Zhang2017gmsd}          && $0.85$ / $0.85$ \cite{kadid10k} && $0.58$ / $-$                 \\
  MS-GMSD \cite{Zhang2017gmsd}          & $0.81$ / $0.81$ \cite{Zhang2017gmsd}          && $0.85$ / $-$                    && $0.59$ / $-$                 \\
  MS-GMSDc \cite{Zhang2017gmsd}         & $0.89$ / $0.89$ \cite{Zhang2017gmsd}          && $0.87$ / $-$                    && $0.59$ / $-$                 \\
  FSIM \cite{zhang2011fsim}             & $0.80$ / $0.80$ \cite{ponomarenko2015image}   && $0.83$ / $0.83$ \cite{kadid10k} && $0.59$ / $0.60$ \cite{pipal} \\
  FSIMc \cite{zhang2011fsim}            & $0.85$ / $0.85$ \cite{ponomarenko2015image}   && $0.85$ / $0.85$ \cite{kadid10k} && $0.59$ / $-$                 \\
  SR-SIM \cite{zhang2012srsim}          & $0.81$ / $0.81$ \cite{athar2019comprehensive} && $0.84$ / $0.84$ \cite{kadid10k} && $0.54$ / $-$                 \\
  SR-SIMc \cite{zhang2012srsim}         & $0.87$ / $-$                                  && $0.87$ / $-$                    && $0.57$ / $-$                 \\
  VSI \cite{zhang2014vsi}               & $0.90$ / $0.90$ \cite{athar2019comprehensive} && $0.88$ / $0.86$ \cite{kadid10k} && $0.54$ / $-$                 \\
  MDSI \cite{nafchi2016mdsi}            & $0.89$ / $0.89$ \cite{nafchi2016mdsi}         && $0.89$ / $0.89$ \cite{kadid10k} && $0.59$ / $-$                 \\
  HaarPSI \cite{Reisenhofer2018haarpsi} & $0.87$ / $0.87$ \cite{Reisenhofer2018haarpsi} && $0.89$ / $0.89$ \cite{kadid10k} && $0.59$ / $-$                 \\
  Content\textsubscript{VGG16} 
  \cite{johnson2016perceptual}          & $0.71$ / $-$                                  && $0.72$ / $-$                    && $0.45$ / $-$                 \\
  Style\textsubscript{VGG16} 
  \cite{johnson2016perceptual}          & $0.54$ / $-$                                  && $0.65$ / $-$                    && $0.34$ / $-$                 \\
  LPIPS\textsubscript{VGG16} 
  \cite{zhang2018unreasonable}          & $0.67$ / $0.67 $ \cite{ding2020image}         && $0.72$ / $-$                    && $0.57$ / $0.58$ \cite{pipal} \\ 
  DISTS \cite{ding2020image}            & $0.81$ / $0.83$ \cite{ding2020image}          && $0.88$ / $-$                    && $0.62$ / $0.66$ \cite{pipal} \\
  PieAPP \cite{prashnani2018pieapp}     & $0.84$ / $0.88$ \cite{ding2020image}          && $0.87$ / $-$                    && $0.70$ / $0.71$ \cite{pipal} \\
  DSS \cite{Balanov2015dss}             & $0.79$ / $0.79$ \cite{athar2019comprehensive} && $0.86$ / $0.86$ \cite{kadid10k} && $0.63$ / $-$                 \\
  \cmidrule{1-6}
  
\multicolumn{3}{c}{No-reference metrics} \\
\cmidrule{1-6}

  BRISQUE \cite{MittalMB12}             & $0.37$ / $0.84$ \cite{athar2019comprehensive} && $0.33$ / $0.53$ \cite{kadid10k} && $0.21$ / $-$                 \\
  
\cmidrule{1-6}

\multicolumn{3}{c}{Distribution-based metrics} \\
\cmidrule{1-6}
  KID\textsubscript{InceptionV3} 
  \cite{binkowski2018demystifying}      & $0.42$ / $-$                                  && $0.66$ / $-$                    && $0.12$ / $-$                 \\
  FID\textsubscript{InceptionV3} 
  \cite{heusel2017gans}                 & $0.67$ / $-$                                  && $0.66$ / $-$                    && $0.18$ / $-$                 \\
  GS\textsubscript{InceptionV3} 
  \cite{khrulkov2018geometry}           & $0.37$ / $-$                                  && $0.37$ / $-$                    && $0.02$ / $-$                 \\
  IS\textsubscript{InceptionV3} 
  \cite{salimans2016improved}           & $0.26$ / $-$                                  && $0.25$ / $-$                    && $0.09$ / $-$                 \\
  MSID\textsubscript{InceptionV3} 
  \cite{tsitsulin2020shape}             & $0.21$ / $-$                                  && $0.32$ / $-$                    && $0.02$ / $-$                 \\

\bottomrule
\end{tabular}
}
\end{center}

\begin{tablenotes}
  \small
  \item \textbf{Note 1}: Typically, the distance between correlation values obtained with new and reference implementations is used to verify the accurateness of the former. 
  Zero difference is an indication of correctness.
  However, we find it peculiar to find that for some metrics (\textit{e.g.}, SSIM, MS-SSIM, VSI) the same implementation exactly matches with only one of two reference values.  
  \item \textbf{Note 2}: All three considered datasets are widely used in the IQA research community to assess metrics on their ability to estimate IQ in the same domain - general natural images. 
  However, we observe a significant drop of SRCC values for all metrics on the larger PIPAL dataset compared to smaller TID2013 and KADID10k datasets.
  While a domain shift caused by larger variety or distortions introduced in the newer PIPAL dataset could explain the observation, more investigations are required.
\end{tablenotes}

\label{table:metrics-comparison}
\end{table*}

\subsection*{IQA datasets}

IQA datasets are typically designed to evaluate existing IQMs on their ability to answer previously not considered questions such as IQ from two different view distances (CID dataset \cite{Liu2014CIDIQ}), IQ on images perturbed with multiple types of distortions (MDID \cite{sun2017mdid}, MD-IVL \cite{corchs2017multidistortion} and VLC \cite{Zaric2011vcl} datasets), evaluation of GAN-based image restoration algorithms (PIPAL \cite{pipal} and NTIRE 2021 \cite{Gu2021ntire} challenges and datasets).
Another commonly used datasets are LIVE \cite{LIVE_MD}, TID2013 \cite{ponomarenko2015image} and KADID-10k \cite{kadid10k}. 
Some works even try to propose a method to construct a general-case IQA dataset with extremely diverse image characteristics \cite{Lin2018kaniq10k}.

\subsection*{Subjective annotations}
The majority of the listed datasets provide image pairs accompanied by subjective quality scores.
Typically, subjective scores are estimated by human assessors and aggregated in the form of Mean Opinion Scores (MOS) or Differential Mean Opinion Scores (DMOS).
In some cases, raw scoring data is first converted to z-scores averaged and re-scaled from 0 to 100 to account for different scoring across respondents as proposed in \cite{Sheikh2006}.

\subsubsection*{Datasets selection}

A significant number of IQA databases have come out over the last 15 years. 
There is currently no \textit{gold standard} dataset. 
Table \ref{table:iqa_datasets} shows that IQA datasets use a variety of subjective testing methodologies, number of images, and number of distortions. 

Typically, usage of datasets with a small number of images and distortions results in high variance between evaluation results. 
Considering that, our main criteria for selection were the size of the dataset and distortions variety.

We selected TID2013 \cite{ponomarenko2015image} and KADID-10k \cite{kadid10k} databases as one of the most popular in the research community and PIPAL \cite{pipal} as the one with a higher variety of introduced distortions. 
After that, we selected an evaluation criterion (type of correlation) that can be computed on selected datasets and compared with the reference values.

\subsection*{Evaluation criteria}

Among numerous evaluation criteria, PLCC, SRCC, and KRCC are the most commonly used in large-scale studies of IQ metrics.

Pearson linear correlation coefficient (PLCC) requires produced scores to be linear with respect to subjective ratings. 
Previous studies \cite{Mason2020medicalcomparison, kastryulin2022iqa_mri} showed non-linear relation between IQM scores and human accessors' scores.
A common solution is to apply a non-linear regression by adopting the five-parameter modified logistic function \cite{sheikh2006statistical}. 
Even though this approach solves the non-linearity problem, we do not use it due to the poor reproducibility of model fitting results among studies.

Spearman's rank-order correlation coefficient (SRCC) and Kendall rank correlation coefficient (KRCC) measure monotonicity between two measured quantities.
Despite different calculation methods, KRCC is found to be highly consistent with the SRCC \cite{athar2019comprehensive}.
Considering that, we use only the most popular SRCC score for further evaluations.

\begin{equation}
\text{SRCC} = 1 -  \frac{6 \sum_{i=1}^{n} d_i ^ 2}{n (n^2 - 1)},
\end{equation}
\noindent
where $d_i$ is the difference between the \textit{i}-th image’s ranks in the objective and the subjective ratings and $n$ is the number of observations.

\begin{figure*}[!ht]
    \centering
    \includegraphics[width=\textwidth]{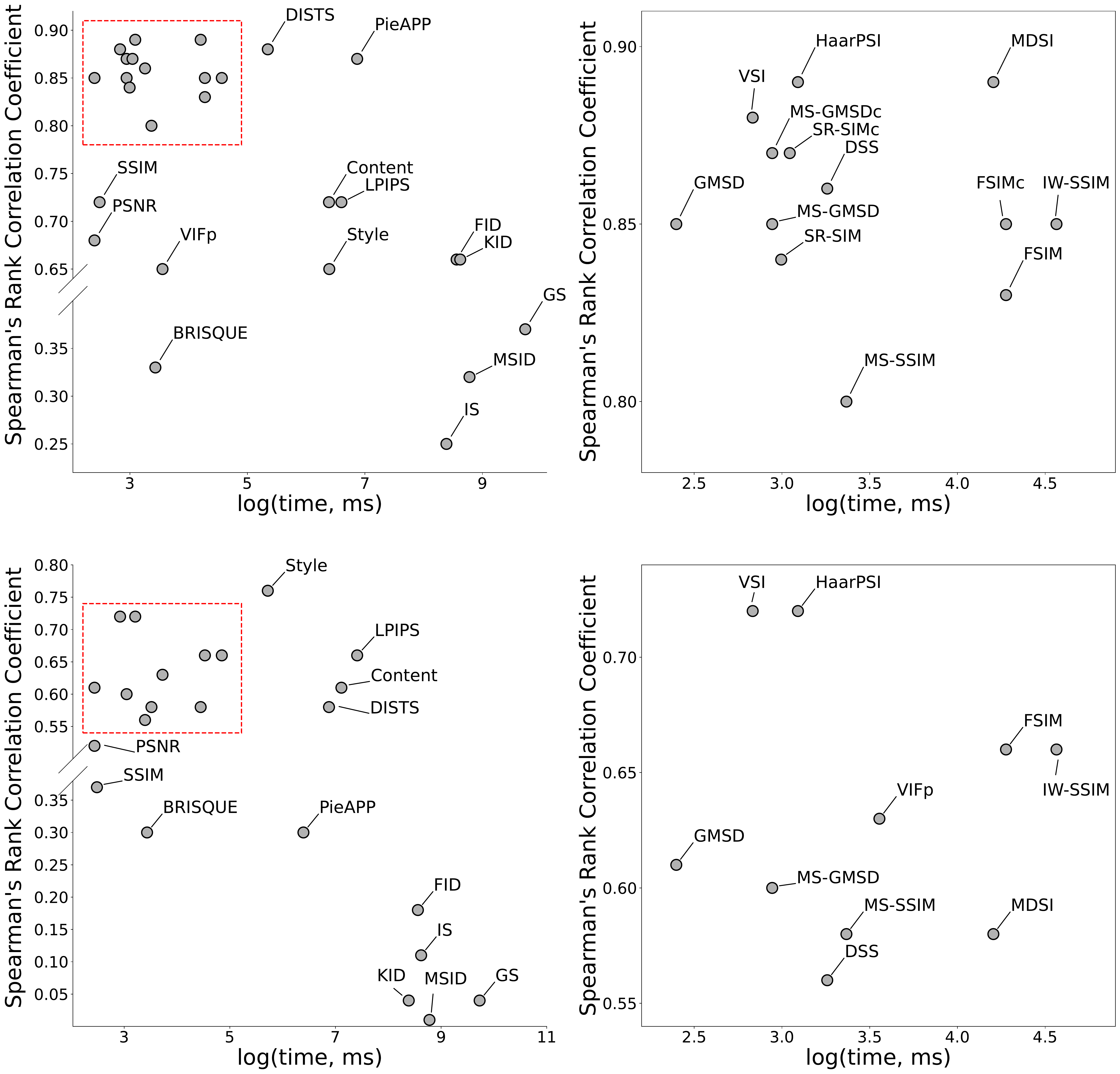}
    \caption{Relationship between the computation time of a metric on \textbf{CPU} and their  performance in terms of SRCC score on Natural Images from KADID-10k dataset \cite{kadid10k} (top row) and MRI Images \cite{kastryulin2022iqa_mri} (bottom row) for all metrics (left column) and zoomed-in region indicated in red (right column). Metrics with the best time-quality relation are located in the top-left corner.}
    \label{fig:complexity-performance-cpu}
\end{figure*}

\begin{figure*}[!ht]
    \centering
    \includegraphics[width=\textwidth]{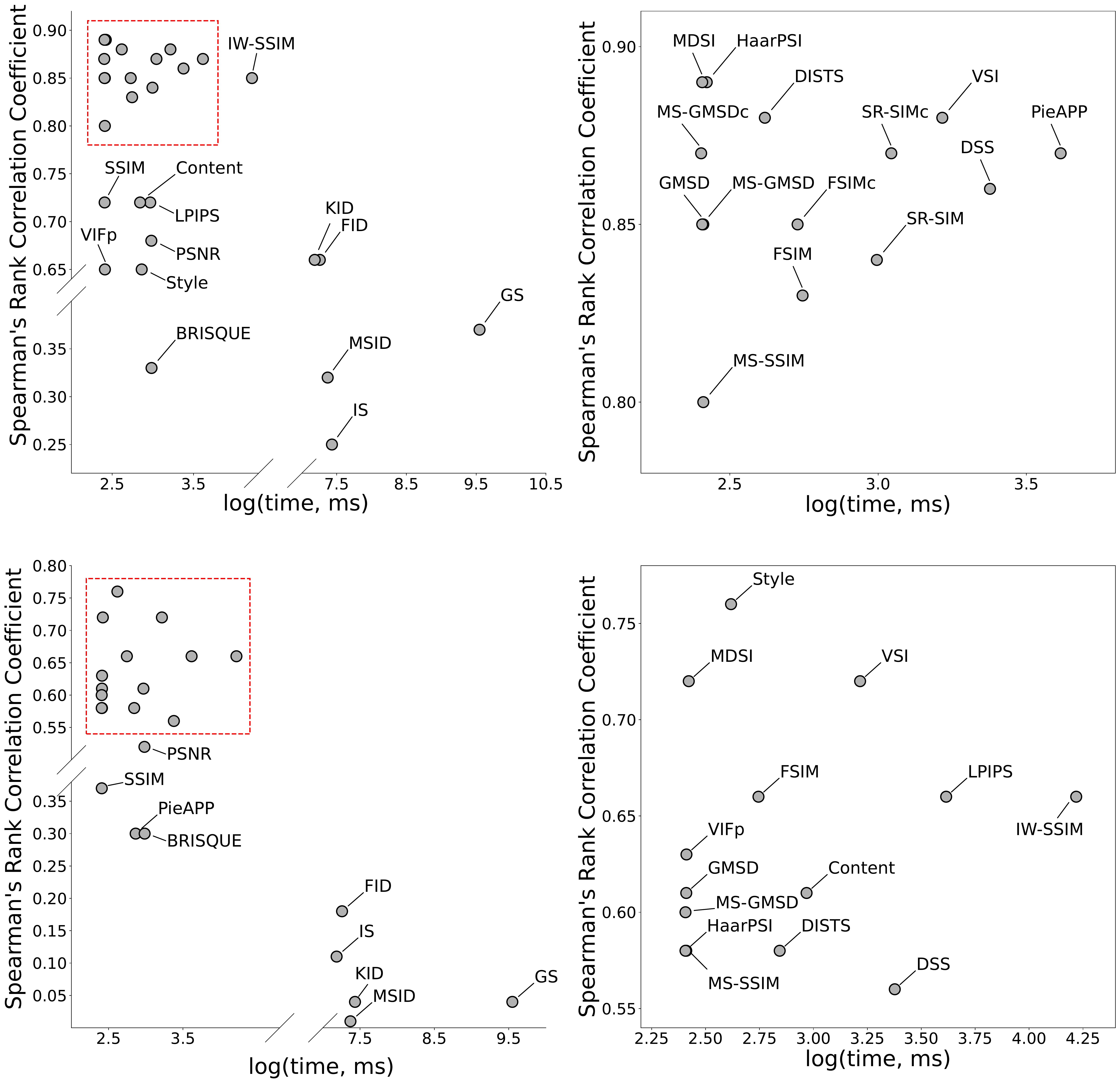}
    \caption{Relationship between metrics' computation time on \textbf{GPU} and their performance in terms of SRCC score on Natural Images from KADID-10k dataset \cite{kadid10k} (top row) and MRI Images \cite{kastryulin2022iqa_mri} (bottom row) for all metrics (left column) and zoomed-in region indicated in red (right column). Metrics with the best time-quality relation are located in the top-left corner.}
    \label{fig:complexity-performance-gpu}
\end{figure*}

\subsection*{Implementation details of DB IQMs}

DB IQMs are not originally designed for pair-wise comparison of images. 
Instead, they are intended to be used to compare distributions of two image sets.
However, we investigate DB metrics for pair-wise comparison of images following the computation strategy proposed in \cite{kastryulin2022iqa_mri}.
To encounter for the initial setting, we represent each image in a pair as a set of overlapping patches of size $96 \times 96$ with \textit{stride} = $32$, which allows us to reformulate the pair-wise comparison of images as a comparison of patch distributions.
After we extract features from two sets of patches, we proceed with the initial flow of metrics' computation.

\subsection*{Confirmation of implementations correctness}

All implementations in the PyTorch Image Quality library are verified to be consistent with the original implementations proposed by the authors of each metric on selected IQA datasets. 
Refer to Table \ref{table:metrics-comparison} for a detailed comparison.

Typically, IQMs are evaluated on a single dataset. 
Our verification results allow comparing metrics performance across different image sources to show that even correct implementations (that match correlation values on initial datasets) may not match values reported on different datasets.
It is also worth mentioning that the performance of some feature extraction-based metrics (\textit{e.g.} BRISQUE) cannot be fully reproduced on IQA datasets even though the code match the official implementation of the metric provided by their authors.
\section{Performance -- complexity trade-off}

Figures \ref{fig:complexity-performance-cpu} and \ref{fig:complexity-performance-gpu} describe performance comparison of the implemented metrics in terms of SRCC and computational time on CPU and GPU for Natural Images from KADID-10k \cite{kadid10k} dataset and MRI Images \cite{kastryulin2022iqa_mri}.
Benchmarks were performed on a dedicated instance of NVIDIA DGX Station with Intel Xeon E5-2698 v4 CPU and four NVIDIA V100 32Gb GPUs. 
A single graphic card was used in all experiments for convenience and simplicity of comparison.
Even though no other processes except from system utilities of the GNU/Linux 5.4.0-91-generic x86-64 operating system were running during the performance of the benchmarks, we recommend to focus on relative performance of metrics rather than the exact numbers of their computation time.

Several key observations can be made based on results of the benchmarks.
DB IQMs tend to group in the lower-right corner, steadily showing poor performance both in terms of CPU and GPU computation time and SRCC values (except from FID\textsubscript{VGG16} on MRI data).
Even though GPU acceleration significantly speeds up their computation, the computation time gap remains to be considerable.

The best trade-off between quality and performance is achieved by metrics located in the upper-left corner. 
In all experiments there is a group of algorithms that tend to group together, forming a category of methods attractive for practical application.
While the composition of this group varies slightly, there are several metrics that consistently perform well regardless of the domain and computing device in question such as MDSI, VSI, HaarPSI, DSS, GMSD and several others. 
Feature-based FR IQMs such as DISTS and PieAPP show high SRCC values on Natural Images but take long to be computed on CPU.
GPU acceleration plays the key role for these metrics, putting them to the lower-left group of top performers.
Widely used PSNR and SSIM are easy to compute both on CPU and GPU but they compromise IQA quality on both considered domains.
\section{Conclusion and future work}

The main contribution of our work is the PyTorch Image Quality (PIQ) Assessment toolbox \cite{piq} with a diverse set of measures, verified according to their formal definitions and the original authors' implementations.
The PIQ package facilitates the performance evaluation of any computer vision solution where an image-to-image task is performed.
In addition, we provide the comparison of common Image Quality Metrics on the image datasets from the general and the medical domains, assessing the SRCC and the computation time.

The future development of the PyTorch Image Quality library will be aimed at supporting the latest trends and advances in the objective Image Quality Assessment and at the improvement of the usability and the scalability of the implemented algorithms.
\section{Acknowledgments}

We greatly appreciate all contributions from the members of the PIQ community, including questions, discussions, design suggestions, and technical implementations.

\bibliographystyle{unsrt}  
\bibliography{main}

\appendix
\newpage\section{Appendix. Quality Metrics}
\label{sec:appendix}

\begin{figure*}[ht!]
\centering
  \includegraphics[width=\textwidth]
    {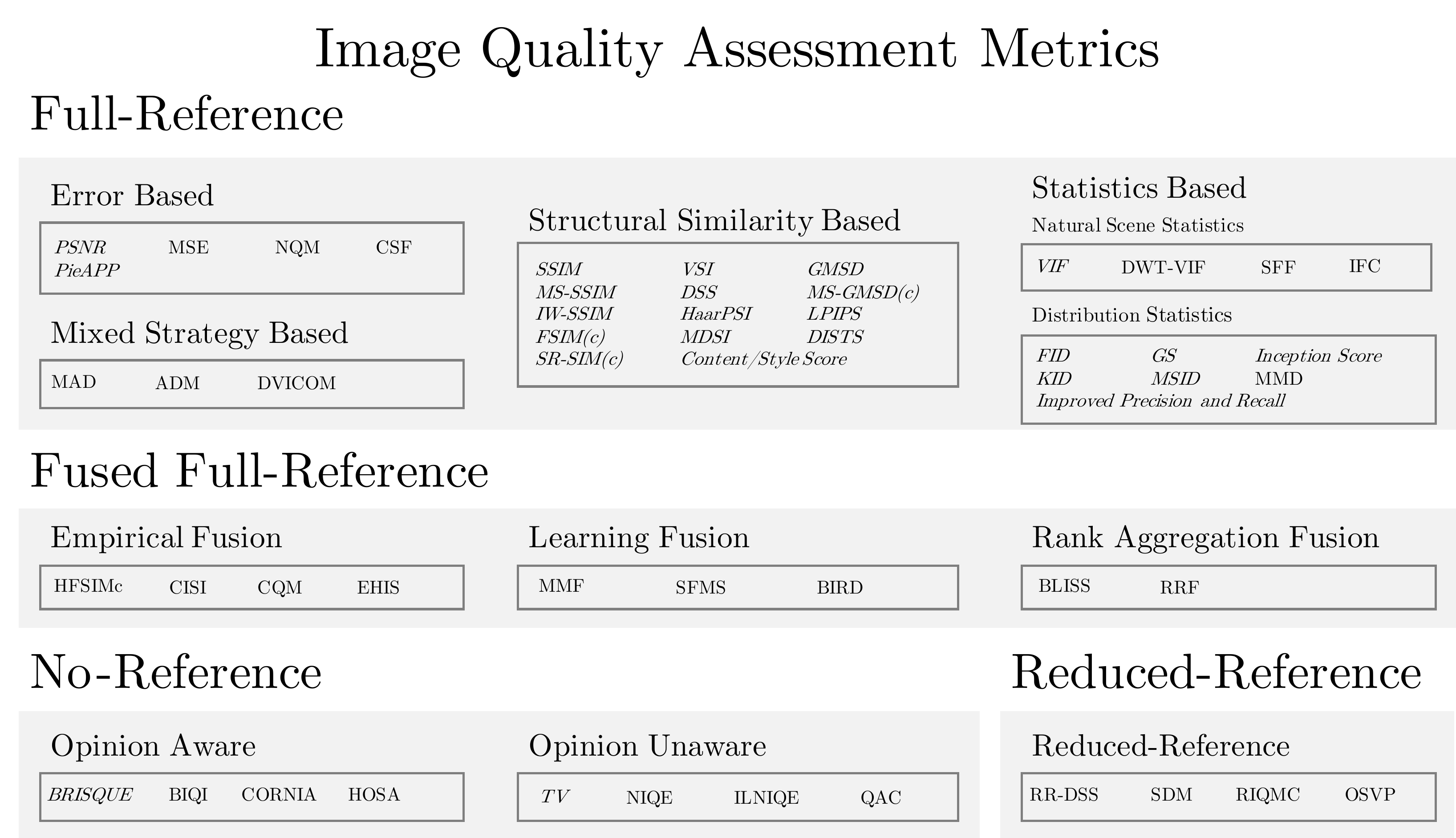}
  \caption{General taxonomy of IQA metrics. Italic metrics are available with PIQ package. ADD WSPNR, IW-PSNR, SWDN.}
  \label{metrics_tree}
\end{figure*}

Historically, the first works on perceptual full-reference IQA appeared almost half a century ago, with the pioneering work of Sakrison and Mannos \cite{mannos1974effects} focusing on a class of visual fidelity criterion in the context of image encoding. 
Over the past few decades, a number of alternative models mimicking certain functionalities of the Human Visual System (HVS) were proposed.

Such approaches couldn't model the real HVS, which is a complex and highly nonlinear system, while most models rely on simplifications and strong assumptions (e.g. linearity or quasi-linearity for visual stimuli) and exhibit shortcomings regarding the definition of visual quality, quantification of suprathreshold distortions, and generalization to natural images \cite{wang2006modern}.
Hence, IQMs can be classified and put into a certain category based on their main computation mechanism as shown in figure \ref{metrics_tree}.



In our work, we choose to evaluate representative methods from those categories. A short description of the design philosophies is given below.



\subsection*{Error Based Methods}

Point-by-point comparisons between pixels or convolution responses (e.g, wavelets, CNNs) is the simplest way of measuring perceptual quality.

\textbf{MSE} \cite{wang2004image}, the Mean Squared Error ($\ell_2$-norm), and closely related \textbf{PSNR} \cite{wang2006modern}, the Peak Signal-to-Noise Ratio, are the most frequently used quality metrics, which are a de-facto standart way of measuring image quality. MSE is easy to use, has clear physical meaning (energy of image distortions), satisfies the Parseval’s theorem and can be used for algorithm optimization leading to a closed-form solutions \cite{wang2006modern, wang2009mean}. 
PSNR is defined as a ratio between the maximum possible power of a signal and the power of corrupting noise that affects the fidelity of signal representation \cite{wang2006modern}.

MSE and PSNR have been repeatedly shown to poorly correlate with human judgements in controlled experiments \cite{ponomarenko2015image, kadid10k}.
The main reason for this is four strong underlying assumptions about visual quality \cite{wang2009mean}: 1. Independence of spatial relationships between samples, 2. Independence of relationship between signal and error, 3. Independence of sign of error samples, 4. Equal importance of all signal samples and errors. 
None of this assumptions hold on in real life and accurately describe human's visual system. 

A number of subsequent papers addressed the weaknesses of PSNR and modified it to better suit for IQA.
\textbf{WSNR} \cite{damera2000image}, the Weighted Signal-to-Noise Ratio, used contrast sensitivity function to approximate HVS and assign different weights to signal and noise components, leading to a linear quality measure.
\textbf{NQM} \cite{damera2000image}, the Noise Quality Measure, also used nonlinear quasi-local processing model of the HVS to accomplish quality assessment.

\textbf{IW-PSNR} \cite{wang2010information}, the Information Weighted PSNR, is a continued idea of applying additional "weights" to address importance of different areas.
It uses information theoretic principles to add weights for regions of visual content, which are perceptually more important than others, either due to the visual attention property of the HVS or due to the influence of distortions.

\textbf{MAD} \cite{larson2010most}, the Most  Apparent  Distortion measure, explicitly models adaptive strategies of the human visual system. 
For high quality images with only near-threshold distortions a detection-based strategy is employed, and an appearance-based strategy is activated if the distortions are clearly visible.
The results are then combined into a single score by weighting scheme, where importance of each strategy is dependent on the distortions strength.



Following the success in image classification \cite{krizhevsky2012imagenet} and recent developments in deep learning \cite{lim2017enhanced, ledig2017photo} usage of convolutional neural networks (CNN) became a popular way for constructing IQA methods.
A double-path CNN was introduced by the pioneering work of Liang et al. \cite{liang2016image}.
Here it was proposed to crop images into small patches and propagate them through a dedicated network branches to obtain patch scores.
The scores determined the final overall image quality score by averaging predicted patch-wise values. 
The model was trained using regression to MOS scores from TID2013 \cite{ponomarenko2015image} database.

Another popular approach, proven to be useful for many image processing tasks, is the usage of generic features obtained from pre-traind networks, such as AlexNet \cite{krizhevsky2012imagenet}, VGG \cite{simonyan2014very}, SqueezeNet \cite{iandola2016squeezenet} and others.
\textbf{Perceptual Loss} and \textbf{Style Score} were introduced by Johnson et al. \cite{johnson2016perceptual}. 
\textbf{Perceptual Loss} used a single-path CNN trained on ImageNet ILSVRC 2012 \cite{krizhevsky2012imagenet} to extract deep feature representations of reference and distorted images. 
Those features were then compared by taking \textbf{MSE} and averaging error between spatial and feature dimensions.
\textbf{Style Score} extracts and compares texture information from the images by computing Gram matrices between features from a CNN and taking \textbf{MSE} between them. 

In \textbf{LPIPS} \cite{zhang2018unreasonable}, the Learned Perceptual Image Patch Similarity, Zhang et al. pointed out that methods based on utilization of deep features significantly outperform traditional algorithms on a wide range of distortions.
LPIPS computes the distance between feature representations on multiple levels, similar to Perceptual Score, building on a premise that different layers represent different structures of the image. 
To proper combine features, they are first unit-normalized to have similar scales and then summed with weights learned on BAPPS \cite{zhang2018unreasonable} dataset to minimize error between model and humans preference over two images. 

\textbf{PieAPP} \cite{prashnani2018pieapp}, the Perceptual Image-Error Assessment through Pairwise Preference metric, uses a pairwise-learning framework to predict the preference of one distorted image over the other. 
\textbf{PieAPP} gets features from the feature-extraction (FE) network and then computes the similarity using a score-computation (SC) network. 
Both FE and SC are trained from scratch targeting humans pairwise-preference between images. 

\textbf{SWDN} \cite{gu2020image}, the Space Warping Difference Network, is specifically designed to handle geometric distortions.
When comparing two image features, a new Space Warping Difference layer takes into account not only pixels on corresponding positions as in all pixel-wise methods, but also looks into a small range around them.

Motivated by the above-mentioned results, a number of other FR-IQA algorithms have been proposed relying on different deep features, pooling strategies and pre-trained CNNs \cite{jinjin2020pipal, ding2020image, gu2020image}.

\subsection*{Structural Similarity Based Methods}


IQMs that use certain properties of HVS can be divided into 2 groups: top-down and bottom-up ones. 
In previous section we described bottom-up approaches to IQA design, which used properties of human visual system to modify error-based MSE and PSNR to better simulate  different components of HVS. 
For example, by adding adaptation to luminance, contrast sensitivity and contrast masking \cite{daly1992visible, chou1995perceptually}.

In contrary, approaches following top-down IQA design try to mimic the functionality of HVS as a whole and do not model it by individual components. 
Structural similarity is a perception-based model that considers image degradation as perceived change in structural information, while also including luminance and contrast masking terms. 
These methods are based on the idea that pixels have strong inter-dependencies especially when they are spatially close.

\textbf{SSIM} \cite{wang2006modern}, the Structural Similarity Index Measure, is the most popular top-down approach which has become a de-facto standard in the field of perceptual image processing (along with PSNR) and has inspired subsequent IQA models based on feature similarity \cite{wang2009mean}. 

The main idea behind SSIM is to split image distortions into structural and not-structural and focus on the first ones, as the latter are less noticeable by HVS. 
The comparison is done between luminance (average pixel intensity), contrast (standard deviation of the local image regions) and structure (cross correlation values between two local image regions) components, which are later combined into a single quality map by averaging of local quality scores. 
The main disadvantage of SSIM is that it takes into account only single image scale, and thus can't adapt to different sets of viewing conditions.



\textbf{MS-SSIM} \cite{wang2003multiscale}, the Multi-scale SSIM, measures SSIM on 5 different scales, computing contrast and similarity on all levels, while measuring luminance only at the final scale. 
Resulting scores than combined through a weighted product using weights adjusted on human dataset of mean opinion scores.

\textbf{IW-SSIM} \cite{wang2010information}, the Information content Weighted SSIM, is an extension of MS-SSIM, that added computation of additional content weights based on information theoretic principles.

After the success of SSIM, a lot of new works searched for effective image features that be used to describe contrast, structural information and textures. 

\textbf{ESSIM} \cite{chen2006edge}, the Edge-based SSIM, uses first order difference operators to compare information between image edges and claimed, that edges are the most important structure information for the HVS.

\textbf{GSM} \cite{gsm_measure}, the Gradient Similarity Measure, computes image gradients in horizontal and vertical directions to use those features as an input to the structural similarity computation. 

    
\textbf{GMSD} \cite{xue2013gradient}, the Gradient Magnitude Similarity Deviation, is focused on computational efficiency of quality predictions. 
Based on the idea that global variation of image local quality degradation can reflect its overall quality, authors proposed to compute the standard deviation of the pixel-wise gradient similarity map as an IQA index. 
This method is, however, problematic because an image with a large but constant local distortion yields a standard deviation of zero, indicating the best predicted quality. 

\textbf{MS-GMSD} \cite{zhang2017gradient}, the Multi-scale GMSD, accounts for the variations in viewing conditions by measuring GMSD on 5 different scales, similarly to MS-SSIM.
The chromatic version, named MS-GMSDc, additionally measures chrominance dissimilarity on the last scale, motivated by lower HSV sensitivity to colour distortions. 

\textbf{FSIM} \cite{zhang2011fsim}, the Feature Similarity Index Measure, is built on assumption that low-level features obtained in the early stage of HVS information processing are used for understanding the image content. 
Two core features used in similarity computations are phase congruency, which is a contrast-invariant dimensionless measure of the local structure, and gradient magnitude maps obtained by Scharr filter. 
During pooling, phase congruency component serves as an adaptive local weighting factor to derive an overall visual quality score.
The color-sensitive version of \textbf{FSIM}, is called \textbf{FSIMc} \cite{zhang2011fsim}.
It first converts RGB images into YIQ color space, and then computes FSIM on luminance channel and chrominance similarity maps on I and Q channels.

    

\textbf{VSI} \cite{zhang2014vsi}, the Visual Saliency Induced quality index, states that the change of salience in degraded areas is the major predictor of image quality. 
The visual saliency index is computed using phase congruency and two simple priors (color temperature and center priors). 
Gradient magnitude maps are used as an additional feature and pooling strategy mimics FSIM measure.
\textbf{VSI} shows a good correlation with human judgments on localized distortions, such as compression artefacts or local patch substitutions. 

Recently proposed \textbf{HaarPSI} \cite{reisenhofer2018haar}, the Haar perceptual similarity index, decomposes both distorted and reference images into Haar wavelets and computes the structural similarity between magnitudes of high-frequency coefficients. 
The last level of Haar wavelet is used to weight the importance of different regions. 

\textbf{MDSI} \cite{nafchi2016mean}, Mean Deviation Similarity Index, combines the gradient similarity, chrominance similarity and deviation pooling (used in GMSD) into a single IQA model. 
MDSI contains 7 configurable parameters that are jointly optimized on a mean opinion scores dataset to provide best correlation results. 

\textbf{DISTS} \cite{ding2020image}, the Deep Image Structure and Texture Similarity measure, is a deep-learning based IQA algorithm that follows LPIPS design principles. 
Image representations are extracted from the pre-trained convolutional neural network (VGG16 \cite{simonyan2014very}) and combined using SSIM-like structure and texture similarity measurements. 
It is sensitive to structural distortions but at the same time robust to texture resampling and modest geometric transformations \cite{ding2020comparison}.

\textbf{DSS}, the DCT Subbands Similarity, aims to measure changes in structural information using sub-bands in the discrete cosine transform (DCT) domain.
DSS extracts features in block-based DCT subbands and measures the distances between corresponding DCT sub-bands.
The final quality index is computed by pooling those distances together with weights.
The main motivation behind quality assessment in the DCT domain is the observation that the statistics of DCT coefficients change with the degree and type of image distortion.

\subsubsection*{Natural Scene Statistics based}

These methods attempt to measure some approximation of the mutual information between the perceived reference and distorted images as an indication of perceptual image quality. 
Statistical modeling of the image source, the distortion process, and the HVS is critical in algorithm development. 

Visual Information Fidelity (\textbf{VIF}) \cite{sheikh2005visual} uses a Gaussian scale mixture to statistically model the wavelet coefficients of a steerable pyramid decomposition of an image \cite{simoncelli1992shiftable}. 
After that, it predicts the distorted image quality by quantifying the amount of preserved information from the reference image. 
Its spatial domain, named VIFp, computes fidelity on raw pixels. 

    

\subsubsection*{Distribution Statistics based}

This class of methods originated from the domain of generative modeling, where evaluation by a direct comparison is not possible and only the distance between model distributions can be measured. 


Freshet Inception Distance (\textbf{FID}) \cite{heusel2017gans} considers embeddings of the two data distributions as a product of continuous multivariate Gaussian. 
The mean and covariance are estimated for both, and the Fréchet distance between these two Gaussians (a.k.a Wasserstein-2 distance)  is then used to quantify the quality of the generated sample.

Kernel Inception Distance (\textbf{KID}) \cite{binkowski2018demystifying}, sometimes referred to as Maximum Mean Discrepancy (MMD) -  computes the dissimilarity between two probability distributions $P$ and $Q$ by measuring squared MMD for some fixed characteristic kernel function (e.g., Gaussian kernel).
    
Multi-Scale Intrinsic Distance (\textbf{MSID}) \cite{tsitsulin2020shape} develops an intrinsic and multi-scale method for characterizing and comparing data manifolds, using a lower-bound of the spectral Gromov-Wasserstein inter-manifold distance, which compares all data moments.
    
Geometry Score (\textbf{GS}) \cite{khrulkov2018geometry} constructs a performance measure by comparing geometrical properties of the underlying data manifolds.
In particular, topological approximation for computation of connected components ("holoes") in homology using witness complex \cite{de2004topological} is introduced.













\subsection*{No-Reference Opinion Aware Methods}

No-Reference (NR) IQA methods evaluate distorted image's quality in the absence of any reference information \cite{sheikh2006statistical}.
Thus they are also referred to as blind IQA methods. 

Blind/Referenceless Image Spatial Quality Evaluator (\textbf{BRISQUE}) operates on locally normalized luminance values in the spatial domain, called Mean Subtracted Contrast Normalized (MSCN) coefficients. 
Various features are extracted from the MSCN coefficients and their pairwise products, which are then used to estimate Generalized Gaussian Distribution (GGD) and  Asymmetric GGD parameters. 
SVR model is then used to learn to map the features to the quality score. 

    
Inception score (\textbf{IS}) \cite{salimans2016improved} uses the pre-trained InceptionNet model for feature extraction and captures properties of generated sample to capture properties of generated samples: diversity with respect to class labels and high classifiability.

Improved Inception score (\textbf{IS'}) \cite{barratt2018note} proposed a different way of feature aggregation, which improved both calculation and interpretability of the Inception Score.


\subsection*{Reduced Reference Methods}

Reduced-reference (RR) image quality assessment methods estimate the quality of distorted images using only partial information about the reference images when the full-reference approaches are unfeasible.
Partial information represented by extracted features should be relevant to the human judgement of image quality and sensitive to a variety of image distortions.
Consequently, the challenge of a reduced reference metric is to find optimal trade-off between the amount of information represented by extracted features and their correlation with HVS.

\textbf{RR-DSS} \cite{balanov2016reduced}, a reduced-reference image quality assessment based on DCT sub-band similarity measure.
It is an adapted full-reference DSS quality assessment measure, which assumes that HVS adapts for extracting structural information.
RR-DSS uses only a few lowest frequency sub-bands for the quality assessment.
In order to maintain good IQA results, \textbf{RR-DSS} should use at least 3 to 10 sub-bands.
RR-DSS uses spatial down-sampling to reduce amount of information about reference image.
As typical distortions are usually spread over the image, uniform down-sampling of the local variances preserves important features of the distortions.
However, down-sampling makes it impossible to compute the cross-correlation for the DC sub-band.
Therefore, local similarity score for the DC sub-band is computed in the same way as for the AC sub-bands.
RR-DSS showed high correlation with subjective results on average and outperformed most other metrics examined in \cite{balanov2016reduced}, both RR and FR, including SSIM and MS-SSIM.
In addition, the method has a simple implementation and incurs low computational complexity, while the trade-off between side information and image quality estimation accuracy can be adjusted according to the task.

\textbf{SDM} \cite{gu2013new}, a reduced-reference IQA using Structural Degradation Model, consists of two main stages: acquiring structural degradation information for distorted and original images and consecutive aggregation into a single score. 
Structural degradation information (SD) can be defined as structural similarity indexes between mean features and variance features obtained using kernels with various variation values computed for different parts of the images.
The SD information is used to derive distances between distorted and reference images, which are linearly combined using regression model parameters optimised by training.
In particular, one can use SVM \cite{chang2011libsvm} for regression as a model and get an \textbf{S-SDM} score.
The SDM approach relies on various spatial responses providing low computational complexity and fast execution.
However, optimisation of model parameters leads to dependency on the dataset used for training, which might limit the generalisation properties of the metric between different image domains.

\textbf{RIQMC} \cite{gu2013subjective}, a reduced-reference image quality metric for contrast-changed images, aims to evaluate image quality according to contrast features of a picture such as skewness and kurtosis of the image histograms, which correlates with HVS \cite{motoyoshi2007image,zoran2009scale}.
RIQMC is based on the first four order statistics, which are evaluated for a distorted image, and values of entropy for distorted and reference images.
The final value of RIQMC is a linear combination of the entropy and the first four order statistics.
Even though the RIQMC metric needs only an entropy of a reference image, it outperforms full-reference metrics such as PSNR, SSIM, MS-SSIM, IW-SSIM, and MAD, on the CID2013 dataset and TID2008 subset according to PLCC and SROCC \cite{gu2013subjective}.
However, the performance of RIQMC may vary for general image distortions as the method was proposed exclusively for contrast-changed images.

\textbf{OSVP} \cite{wu2016orientation}, an orientation selectivity-based visual pattern IQA, is a reduced-reference measure inspired by the orientation selectivity (OS) mechanism for visual content extraction by the human visual system.
When visual content is observed, the input signal interacts with the visual cortex depending on the spatial arrangement in local perceptual fields generating OS visual patterns.
In order to represent the OS mechanism, gradient directions are extracted per pixel, building the spatial relationship between each pixel and its local neighbours (OSVP).
Next, the content of an image is mapped into a histogram according to spatial relationship patterns.
The final score is calculated as changes between the histograms for reference and distorted images.
Being inspired by neuroscience findings in orientation selectivity mechanism, OSVP RR-IQA showed performance consistent with HVS perception on five publicly available databases with limited reference data.

\newpage\section{Appendix. Additional Benchmarks}

\begin{figure*}[!ht]
    \centering
    \includegraphics[width=\textwidth]{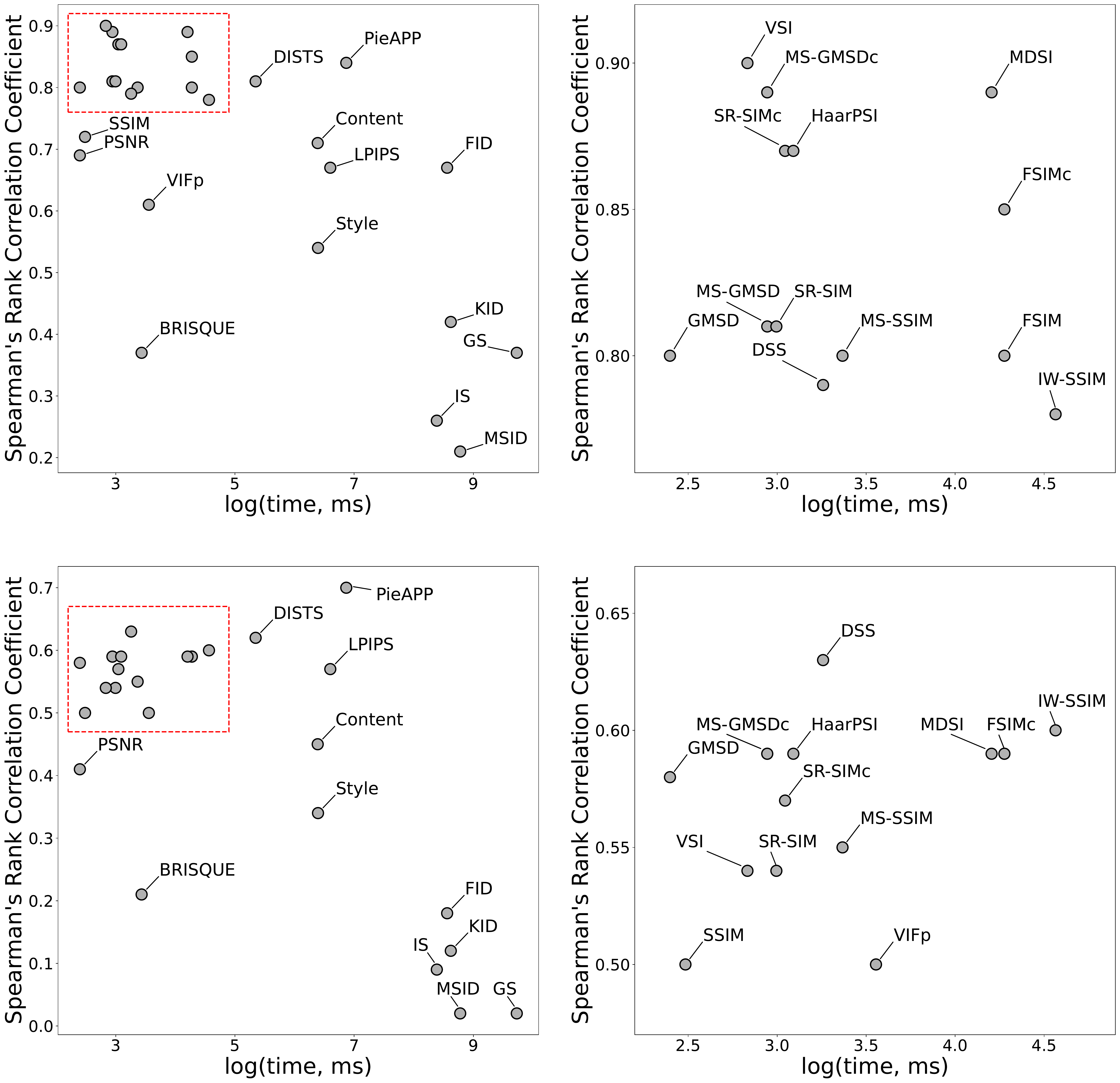}
    \caption{Relationship between metrics' computation time on \textbf{CPU} and their  performance in terms of SRCC score on Natural Images from TID2013 \cite{ponomarenko2015image} (top row) and PIPAL \cite{pipal} (bottom row) datasets for all metrics (left column) and zoomed-in region indicated in red (right column). 
    Metrics with the best time-quality relation are located in the top-left corner.}
    \label{fig:complexity-performance-cpu-tid-pipal}
\end{figure*}

\begin{figure*}[!ht]
    \centering
    \includegraphics[width=\textwidth]{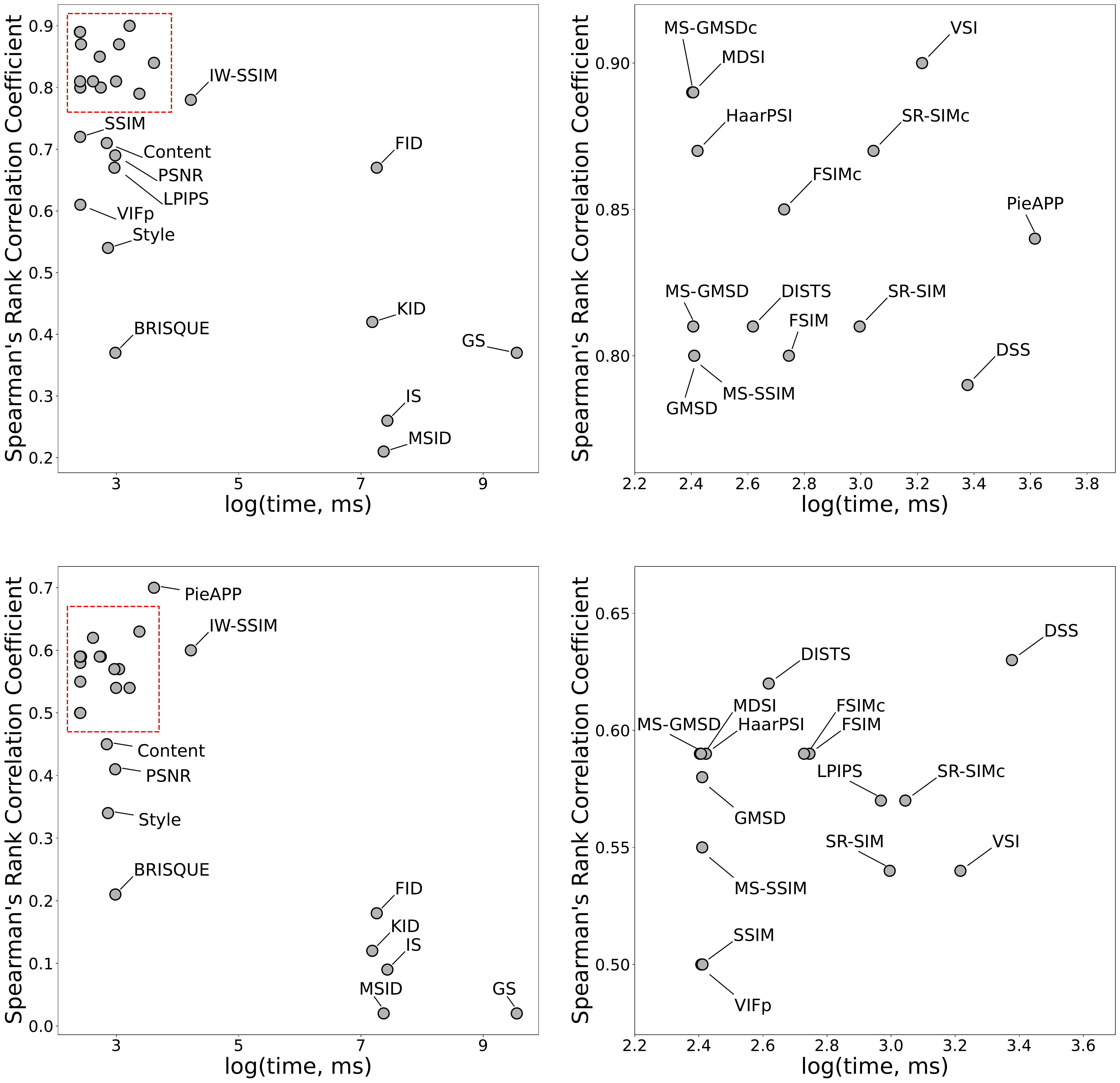}
    \caption{Relationship between metrics' computation time on \textbf{GPU} and their  performance in terms of SRCC score on Natural Images from TID2013 \cite{ponomarenko2015image} (top row) and PIPAL \cite{pipal} (bottom row) datasets for all metrics (left column) and zoomed-in region indicated in red (right column). 
    Metrics with the best time-quality relation are located in the top-left corner.}
    \label{fig:complexity-performance-gpu-tid-pipal}
\end{figure*}

The main part of the work considered the  performance -- complexity trade-off for KADID10k \cite{kadid10k} and MRI reconstructions \cite{kastryulin2022iqa_mri} datasets. 
Here we present additional benchmarks for two commonly used datasets: TID2013 \cite{ponomarenko2015image} and PIPAL \cite{pipal} of CPU (Fig. \ref{fig:complexity-performance-cpu-tid-pipal}) and GPU (Fig. \ref{fig:complexity-performance-gpu-tid-pipal}).

\end{document}